\documentclass[prl,twocolumn,amsmath,amssymb,superscriptaddress,showpacs]{revtex4}
\usepackage{graphicx}
\usepackage{dcolumn}


\newcommand{\la}{\label}
\newcommand{\be}{\begin{equation}}
\newcommand{\ee}{\end{equation}}
\newcommand{\bea}{\begin{eqnarray}}
\newcommand{\eea}{\end{eqnarray}}
\newcommand{\p}{\partial}

\newlength{\intwidth}
\DeclareRobustCommand{\fpint}[2]
   {\mathop{%
      \text{%
        \settowidth{\intwidth}{$\int$}%
        \makebox[0pt][l]{\makebox[\intwidth]{$-$}}%
        $\int_{#1}^{#2}$}}}

\begin{document}

\title{Quantum Shock Waves -\\
the case for non-linear effects in dynamics of electronic liquids}

\author{E. Bettelheim}
\affiliation{James Frank Institute, University of Chicago, 5640 S.
Ellis Ave. Chicago IL 60637}
\author{A. G. Abanov}
\affiliation{Department of Physics and Astronomy, Stony Brook
University,  Stony Brook, NY 11794-3800}
\author{P. Wiegmann}
\affiliation{James Frank Institute, University of Chicago, 5640 S.
Ellis Ave. Chicago IL 60637} \altaffiliation[Also at ]{Landau
Institute of Theoretical Physics.}

\begin{abstract}
Using  the Calogero model as an example, we show that the
transport in interacting non-dissipative electronic systems is
essentially non-linear. Non-linear effects are due to the
curvature of the  electronic spectrum near the Fermi energy. As is
typical for non-linear  systems, propagating wave packets are
unstable. At finite time shock wave singularities develop, the
wave packet collapses, and oscillatory features arise. They evolve
into regularly structured  localized pulses carrying a
fractionally quantized charge - {\it soliton trains}. We briefly
discuss perspectives of observation of Quantum Shock Waves in edge
states of Fractional Quantum Hall Effect and a direct measurement
of the fractional charge.
\end{abstract}
\date{\today}
\maketitle

\paragraph{1.}
It is commonly assumed that a small and smooth perturbation of
electronic systems remains small and smooth in the course of
evolution. This assumption justifies the linear response theory to
electronic transport.  Although it  is true for dissipative
electronic systems, like a metal, it fails for dissipationless
quantum fluids.

In this paper we argue that in such electronic systems, commonly
known in one spatial dimension, the transport is essentially
non-linear and is subject to an unstable singular behavior.
An arbitrarily small and smooth disturbance of electronic density
and momentum (a wave packet) will inevitably develop to
shot-noise-like oscillatory features, progressively evolving  to
regular localized pulses, carrying quantized charge - a soliton
train. This process is a dissipationless counterpart of a shock
wave, where an initially smooth profile of density tends to
overhang and collapse at some {\it finite} time. A typical
behavior is shown in Fig.~\ref{shocks}. We will demonstrate this
phenomenon on the example of the Calogero model, but believe that
a shock wave instability and the main features of the subsequent
dynamics are universal for dissipationless  interacting electronic
systems. They are insensitive to details of interactions and to
the initial profile of the wave.

We note that, for some physical systems, the time of the onset of
the shock wave may be longer than the time needed for other
physical mechanisms to dissipate the disturbance.  However, our
estimates indicate that the observation of  shock waves at
confined states of a quantum well heterostructures (the edge
states of Fractional Quantum Hall Effect (FQHE) )  is possible. In
addition, a dynamic experiment may allow direct observation of the
fractional charge of ``fly-out'' excitations.

\begin{figure}[b]
    \bigskip
    \begin{center}
    \vspace{-0.5cm}
    \includegraphics[width=8cm]{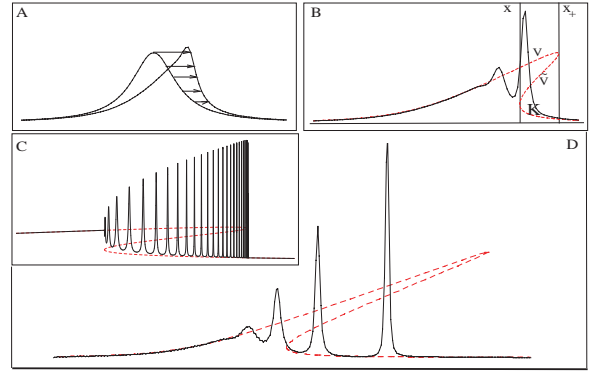}
    \end{center}
\caption{Numerical solution of BO  equation (\ref{BO}). A: The
initial Lorentzian profile containing 7 solitons
is shown
at time
$t=0.9t_{c}$. B: The wave
after a shock is shown  together with the overhanging solution of
Riemann equation (\ref{H}) (red, dashed) at $t= 2.7t_{c}$.
Vertical lines mark the trailing and the leading edges. The moduli
$v, \tilde  v, K$ of the 1-phase  solution (\ref{phase}) are
assigned to different branches of the multi-valued solution of
(\ref{H}).  D: The solutions of BO and Riemann's equations are
shown at $t=11t_{c}$.  C:  Whitham modulated wave represents an
approximate soliton train behavior for an initial Lorentzian shape
of a  larger area.} \la{shocks}
\end{figure}

\paragraph {2. Calogero-Sutherland model} defined on a circle by
\begin{equation}
 \label{1}
    {\cal H} =-\sum_{j=1}^{N} \frac{\hbar^2}{2m}\frac{\p^2}{\p x_j^2}
    +\sum_{k\neq j=1}^{N}
    \frac{\pi^2 }{2L^2}\frac{g}{\sin^2(\frac{\pi}{L}(x_{j}-x_{k}))}
\end{equation}
occupies a special place in 1D quantum physics. The singular
interaction is so strong that it  changes the statistics of
particles from $1$ (Fermi-statistics) to $\lambda$, while
excitations carry statistics of $1/\lambda$  (for a review see
e.g., \cite{1999-Polychronakos,Ha}). The chiral sector of Calogero
model, therefore,
 describes the
edge of the FQHE with a fractional filling $1/\lambda$. Here
$\lambda$ is a dimensionless coupling constant  $g =
\frac{\hbar^{2}} {m}\lambda (\lambda-1) $.  Below we set $a =
(\lambda-1) / \lambda $ , $ \kappa=2 \pi \lambda \hbar/m$, and
rely on results of \cite{AW-2005}.

\paragraph{ 3. Evolution of a semiclassical state.}
Consider an evolution of an initial state - a wave-packet
$|\Psi\rangle$, of a ``large'' size $l\gg k_F^{-1}$  as on
Fig.~\ref{shocks}A. The state is assumed to be semiclassical,
i.e., it involves many particles,  and is such that its Wigner's
function $W(x,p)=\langle\Psi|  e^{\frac{i}{\hbar}(\hat Px +p \hat
X)} |\Psi\rangle$ has a well-defined classical limit.
This means that $W(x,p)$ exponentially vanishes outside of  a
certain star-like support in phase space. Semiclassical
hydrodynamic modes can be viewed as area-preserving waves of the
boundary  of the support. We study the time evolution of the
density $\rho(x,t)=\int W(x,p,t) dp $ and current $j(x,t)=\int
pW(x,p,t) dp $, or velocity $v(x,t)$ defined as $j=m\rho v$ of the
wave-packet. The state can be created by some classical instrument
acting on a system.

First, we argue that in interacting 1D system like (\ref{1}) the
evolution of a semiclassical  state is determined only by
hydrodynamic variables -- the density $\rho(x,t)$ and velocity
$v(x,t)$.  In other words, the semiclassical density and velocity
obey a set of closed evolution equations of hydrodynamic type.
Things are different for non-interacting particles $\lambda=1$.
There,  a ``gradient catastrophe'' also occurs   at a finite time,
but  further evolution has an essentially quantum nature.

\paragraph{4. Quantum hydrodynamics.}
To study the dynamics  of (\ref{1}) and similar problems,  it is
productive to represent a quantum many body problem as quantum
hydrodynamics, i.e., rewrite it solely in terms of  the density
$\rho(x)\!=\!\sum_i\delta(x-x_i)$ and the velocity operators $
\left[\rho(x),v(y)\right]\! = \! - \! i \frac{ \hbar }{ m }
\delta'( x - y ) $  \cite{f}. In this approach \cite{Sakita-book}
excitations appear  as solitons of nonlinear fields $\rho(x)$ and
$v(x)$. Calogero model has been  successfully written in terms of
hydrodynamic  variables, thus extending the familiar
``bosonization'' of fermionic models in 1D
\cite{Jevicki-1992,Awata,AJL-1983,AW-2005}. Below we remind the
major facts following \cite{AW-2005}.

Equations of motion of (\ref{1}) are equivalent to equations of
quantum hydrodynamics
\begin{eqnarray}
   \la{5}
    {\rm continuity\,equation}:&\qquad\dot \rho+\nabla(\rho v)
    &= 0,\\
   \la{6}
    {\rm Euler\, equation}:&\qquad\dot v+\nabla (\frac{v^2}{2} +w)
    &= 0,
\end{eqnarray}
where $\quad w=(\kappa/2\pi )^2 \delta( \rho \epsilon ) / \delta
\rho$ is the quantum enthalpy and $\epsilon$ is the energy per
particle in units of $\lambda\kappa/2\pi$
\begin{equation}\la{7}
    \epsilon= \frac{1}{6}(\pi \rho)^{2}
    +\frac{ a^{2}}{8}(\nabla\ln\rho)^{2}
    +\frac{1}{2}\pi a \nabla\rho_{H},
\end{equation}
and
$\rho_{H}=\fpint{0}{L}\frac{dx'}{L}\rho(x')\cot\frac{\pi}{L}(x'-x)$
is a Hilbert transform of the density. The first term in (\ref{7})
is present already for free Fermi gas ($\lambda=1$) -
the
 Fermi energy. The last two terms absent at $\lambda=1$
reflect the interaction.

\paragraph{5. Sound waves, nonlinear effects and dispersion.}
In the  limit of small  deviations  of density $\delta
\rho\!=\!\rho\!-\!\rho_0\!\ll\!{\rho_0} $, and velocity $\delta
v\!=\!v\!-\!v_0\!\ll\! v_0$ from their mean values $\rho_0$ and
$v_0$, eqs.~(\ref{5},\ref{6}) may be linearized. If in addition
gradients are small $a\frac{\nabla \delta \rho}{ \delta\rho  }\ll
\rho_{0}$ so that the two last terms in (\ref{7}) can be dropped,
one obtains the familiar ``linear bosonization'', where density
and velocity waves move to the right (left) with Fermi velocities
$v_F=v_0 \pm  \frac{\kappa}{2}  \rho_0$ without dispersion.
Introducing left and right moving fields $u=\delta v \pm
\frac{\kappa }{2} \delta \rho$ the
linear hydrodynamics reads
 $\dot{u} + v_s \nabla  u = 0$, where
$v_s = v_F$ are  velocities of sound.

Shifting to a Galilean frame moving with velocity $v_s$, one
recovers non-linear effects:
\begin{equation}
 \label{H}
    \dot u+u\nabla u=0.
\end{equation}
This is is the quantum Riemann equation for 1D compressible
hydrodynamics \cite{Landau_book}. Its follows  from the Galilean
invariance and
 may be traced to  a curvature of the
particle spectrum. The Riemann equation  states that the wave
$u(x,t)$ moves with a velocity which is itself given by $u(x,t)$.
Contrary to the case of a linear spectrum,  when a packet  moves
with the velocity of sound, and does not change its shape, an
arbitrarily small dispersion deforms the packet pushing its denser
part forward.

Eq. (\ref{H}) is equivalent to (\ref{5},\ref{6}) under assumption
of small gradients $a \nabla \delta \rho \ll \delta \rho^{2}$ when
the two last terms in (\ref{7}) are small. However this assumption
almost never holds. It fails for any small and smooth initial
condition with a ``bright side''  when $u(x) \nabla u(x)<0$
(Fig.~\ref{shocks}A). This is  seen from the implicit solution of
(\ref{H}) due to Riemann: $u=f(x-u\, t)$, where $f(x)=u(x,0)$ is
the initial state. The bright side of the wave packet steepens and
finally overturns at some time, $t_c$, where the formal solution
becomes multi-valued \cite{Landau_book}, and un-physical
Fig.~\ref{shocks}B-D.


Hydrodynamics of the electronic system can be understood as a
hydrodynamics of the Fermi sea drawn in the phase space (or more precisely of the support of the Wigner's function in the phase space).  In equilibrium it is bounded  by two lines $k_F^{(\pm)}=m(v_0 \pm \frac{\kappa}{2} \rho_0)$. The field $m u(x,t)$  is interpreted as a wave propagating on the Fermi surface in the phase space.

The time $t_{c}$ is roughly the time for the top of the
wave-packet to travel a distance of its size $l$.  Since the
excess velocity is $u$, an overhang occurs at a time $t_c\!\sim\!
l/(\delta v \pm \frac{\kappa }{2} \delta \rho )$. At this time a
curvature of the spectrum becomes a dominant factor. Obviously, a
smooth and small packet remains in the linear regime longer, but
its life is  finite. In the vicinity of the overhang the classical
Riemann equation becomes invalid. A shock must be regularized
either by quantum corrections at $\lambda=1$, or, in the
interacting case, by the neglected gradient terms.

\paragraph{ \it 6. Dispersive regularization - the role of interaction.}
The role of dispersion changes in the presence of an interaction.
The latter gives  higher gradient corrections to the Euler
equation. In the linear regime they are small, but in the
shock-wave regime, they stabilize growing gradients. This
mechanism is called dispersive regularization, and is well known
in the theory of non-linear waves \cite{book}.  A  stabilization
occurs when dispersive and nonlinear terms in (\ref{7}) become of
same order
\begin{equation}
 \la{cc}
    a \nabla \delta \rho \sim \delta \rho^{2}.
\end{equation}
This condition determines the smallest scale of  oscillations
occurring after a shock wave. It is $\Delta l \sim a /\delta
\rho$.

This is an important result. Rising gradients are bounded by a
scale  exceeding the Fermi length $k_F\Delta l\!\gg\!1$.

As a  consequence {\it a semiclassical wave packet of interacting
1D-fermions remains semiclassical even after entering a shock wave
regime. Its evolution is described by  a classical non-linear
hydrodynamics}.

Condition (\ref{cc}) emphasizes the role of interaction. In its
absence ($\lambda=1$) the   gradient catastrophe is  stabilized
only by quantum effects and  makes a semiclassical description
non-valid at $t\sim t_c$ \footnote{Wave packet in a Fermi gas also
undergoes a shock wave collapse, evolving into Airy type of
oscillations in the growing interval of $x\sim
((\kappa/\rho_0)(t-t_c))^{1/3}$.}.

\paragraph {\it 7. Chiral case} A generic wave-packet  will be
separated into two parts - left  and right modes moving away with
sound velocities. The physics of shock waves is featured in  each
chiral part, so we can treat them separately. Moreover, specifics
of the Calogero model allows us to separate the chiral sectors
exactly, by choosing initial conditions as ${\rm v} = \frac{\kappa
}{2} \rho,$ so that the motion is only (right) chiral. Here ${\rm
v}\!=\!v\!-\! a\frac{\kappa }{4 \pi} \nabla(\log\rho)_H$ is the
shifted velocity  Hermitian with respect to the inner product
\cite{f}. Under this condition two eqs. (\ref{5},\ref{6}) become
identical
\begin{equation}
 \label{ch}
    \dot \rho + \frac{\kappa}{2} \nabla( \rho^2+ \frac{a}{2\pi}\rho\nabla(\log\rho)_H)=0.
\end{equation}

\paragraph{8. Benjamin-Ono equation.}
Equation (\ref{ch}) admits further simplification relevant for the
physics of shock waves. The semiclassical nature of the initial
wave packet insures that deviations of density from the mean
value, $\rho_0$ are small.  Let us choose the chiral case, and a
frame moving with the velocity $v_s = \kappa \rho_0$. Then keeping
gradient  terms in (\ref{ch}) of the lowest order, we linearize
the dispersion term $\rho\nabla(\log\rho)_H\!\approx\!
\nabla(\delta\rho)_H$. As a result we obtain the celebrated
Benjamin-Ono (BO) equation
\footnote{Remarkably,
the physical approximation $\delta\rho\ll\rho_0$ preserves the
integrable structure. A formal expansion in $\rho_0^{-1}$ is, in
fact,  a B\"aklund transformation \cite{paper5}.}
\begin{equation}
 \label{BO}
    \dot u+ u\nabla u+ a\frac{\kappa }{4 \pi }\nabla^2u_H
    = 0 , \quad u= \kappa \delta \rho.
\end{equation}
In hydrodynamics it describes interface waves in a deep stratified
fluid \cite{AC}. Eqs. (\ref{5},\ref{6}) having  a similar
structure but capturing both chiral sectors, were called
Benjamin-Ono equation on the double (DBO) in \cite{AW-2005}. Both
eqs. (\ref{ch},\ref{BO}) are quantum.  The field $u$ in the
quantum BO eq. is the gradient  of a canonical real Bose field.

\paragraph{9. Quantum Benjamin-Ono equation and FQHE.} It is
commonly accepted that edge states in FQHE are described by chiral
bosons \cite{2003-Chang}. The quantum BO equation is a good
candidate for an extension of  the theory of edge states beyond
the linear approximation. Indeed, it has all the necessary
features of the physics of edge states: in the linear
approximation it reduces to linear chiral bosons, the statistics
of excitations is fractional, and it  is Galilean invariant.
Elsewhere we argue that the Galilean invariance and statistics
uniquely determine both the nonlinear and  the dispersion terms of
the equation.

{\it 10. Semiclassical hydrodynamic equations, solitons.} Below we
study a semiclassical version of  quantum hydrodynamics. Due to a
non-linear nature of quantum  equations, the semiclassical limit
is more subtle than merely treating quantum operators as classical
fields. In \cite{AW-2005} we argued that the semiclassical limit
describes a collective motion of excitations (quasi-holes)  with
charge $1/\lambda$. This amounts to replacing the parameter $a$ in
(\ref{5},\ref{6},\ref{ch},\ref{BO}) by the charge of excitation
${1}/{\lambda}$ \footnote{Alternatively, the flow of particles is
described  by the classical hydrodynamics with $a=1$. If the
parameter $\lambda$ is a rational number, the Calogero model has a
rich spectrum of excitations \cite{Ha} sensitive  to the
arithmetic structure of $\lambda$. A semiclassical description of
flows of each type of excitations corresponds to a proper choice
of parameter $a$. } . To see this, one may notice that the
parameter $a$ is a charge of solitons of the classical non-linear
field $u$.

Solitons and more general multiphase solutions of BO are known
\cite{1979-SatsumaIshimori,k}. We have found  general multiphase
solutions and their Whitham modulations for more general  DBO
eqs.(\ref{5},\ref{6}). We present these results elsewhere
\cite{paper5}. Here we restrict ourselves to a simplified
description of a shock wave in the most physically relevant limit
when  BO (\ref{BO}) holds. To this end we need to know only
1-phase and 1-soliton solutions.

The one-phase solution (compare to \cite{Polychronakos-1995})
reads
\begin{equation}
 \label{phase}
     \delta \rho - \overline{\delta \rho}\! =  \!\frac{ 1}{ \pi \lambda} {\rm Im}
    \p_x \log \Big(1\!-\!\sqrt{ \frac{ v \! - \! K  }{\tilde v \!- \! K} }e^{
     \frac{i}{\hbar}\theta(x,t) ) }\Big),
\end{equation}
where $\theta(x,t)= k(x-V t)$,  $k=m( v-\tilde  v)$,
$V=\frac{1}{2}( v+\tilde  v)$ and $\overline{\delta\rho}=\frac{2}{
\kappa}(K+\frac{k}{m})$ are the phase, the wave number, the
velocity and the mean density. Parameters $ v, \tilde  v$ and $K$
are the moduli of the solution.

The  1-soliton solution appears as a long wave limit $k\!\to\!0$
of (\ref{phase})
\begin{equation}
    \delta \rho-\overline{\delta\rho}=\frac{1}{\pi\lambda}{\rm Im} \frac{1}{ x-Vt - i
    \frac{\hbar}{2 m (V-\frac{\kappa}{2}\overline{\delta\rho})}}.
\end{equation}
The amplitude, $\frac{4
(V-\frac{\kappa}{2}\overline{\delta\rho})}{\kappa}$, and the
width, $\frac{
\hbar}{2m(V-\frac{\kappa}{2}\overline{\delta\rho})}$ of the
soliton are determined by the excess velocity $V -
\frac{\kappa}{2} \overline{\delta\rho}$. The charge of the soliton
is $1/\lambda$.

\paragraph {\it 11.  Whitham modulation.}
Although eqs. (\ref{5},\ref{6},\ref{ch},\ref{BO}) are integrable,
only quasi-periodic solutions enjoy  explicit formulas
\cite{paper5}. Solutions with generic initial data can not be
found explicitly. However, in many physically motivated cases, the
solutions are well-approximated by slowly modulated waves. There
it is assumed that the moduli of a quasi-periodic solution (in the
case of 1-phase solution (\ref{phase}) the moduli are $v, \tilde
v, K$) also depend on space and time but in a slow manner. They
do not change much during a period of oscillation. If the scales
of oscillations and modulations are separated, the Whitham theory
provides equations for the space-time dependence of the moduli
\cite {Whitham}.

Shock waves is the most spectacular application of the Whitham
theory. It has been  developed in a seminal paper \cite{GP}  on
the example of the  KdV equation. The idea of the method is the
following.  Outside the shock wave regime a smooth initial wave
remains smooth, and one legitimately neglects the dispersion term,
keeping only  non-linear terms. The equation obtained in this
manner is the Riemann eq. (\ref{H}). Its solution with initial
condition $u(x,0)=f(x)$ reads $u^{(0)}(x,t)=f(x-u^{(0)}(x,t)\cdot
t)$. By using the superscript we indicate that this is not the
oscillatory part of the solution. A solution of (\ref{H}) always
overturns, and, typically becomes a three-valued function in the
interval $x_-(t)<x<x_+(t)$ (Fig.\ref{shocks} B), where the leading
and trailing edges $x_\pm(t)$ are determined by the condition
$\p_x u^{(0)}=\infty$. Let us order the branches as
$u^{(0)}_1>u^{(0)}_2>u^{(0)}_3$.

In the three-valued region the dispersion is important. It
replaces a non-physical overhang by modulated oscillations. The
latter are given by the Whitham's modulated solution. To leading
order one chooses a modulated 1-phase solution. We call it
$u^{(1)}(x,t)$ to emphasize that this solution has only one fast
harmonic. It is glued with $u^{(0)}$ at $x_\pm(t)$:
$u^{(0)}_1(x_-)=u^{(1)}(x_-), \,u^{(0)}_3(x_+)=u^{(1)}(x_+)$.

A modulated 1-phase solution is given  by the formula
(\ref{phase}), where three moduli $ v,\,\tilde  v,\,K$ are smooth
functions of space-time, and  the phase $\theta(x,t)$  is replaced
by a modulated phase $\Theta(x,t)$ found from the Whitham
equations
$$
    \nabla \Theta = m ( v - \tilde  v),\quad \dot \Theta = \frac{m }{2} ( v^2-\tilde  v^2) .
$$
The Whitham equations for the moduli  have again the form of
Riemann equation (\ref{H})
$$
    \dot  v +  v \nabla  v = 0, \;\; \;\,\,\,\dot{ \tilde  v} + \tilde  v
    \nabla \tilde  v=0,\;\;\;\,\,\,\dot K + K \nabla K = 0.
$$
The boundary data for the moduli are chosen such that $u^{(1)}$
stops to oscillate at the gluing points $x_\pm$. That happens  (i)
when $ v \to \tilde  v$ at $x=x_+(t)$, there $u^{(1)} \to K$, and
(ii) when $\tilde  v \to K$, at $x=x_-(t)$, there $u^{(1)}\to
\tilde  v$. The gluing conditions lead to an especially simple
result for the moduli  of BO equation \cite{k,Matsuno}:
$$
    v = u^{(0)}_1(x,t), \; \tilde  v = u^{(0)}_2(x,t), \; K = u^{(0)}_3(x,t) .
$$

\paragraph{\it 12. Shock waves.}
The space-time dependence of the moduli approximately determines
the entire evolution of a wave-packet. We summarize some features
which are not sensitive to its initial shape. Let us choose a
frame moving with the sound velocity. An initially smooth bump
with a height $\delta\rho$ and a width $l$ tends to overhang
toward its bright side. Its dark side becomes  smoother. At the
time $t_c\sim l /\kappa\delta\rho$ the bump starts to produce
oscillations. The oscillations fill the growing interval
$x_-(t)\!<\!x\!<\!x_+(t)$. Further, at $x>x_+(t)$ the shape of the
bump does not change.

The leading edge  runs with an excess  velocity with  respect to
the sound. At large time $x_+\sim \kappa \delta \rho \cdot t$.

The trailing edge $x_-$ slows down  with time. If initial bump
decays as $\delta\rho(x,0) \sim x^{-n}$, then at large time
$x_-(t) \sim t^{1/(n+1)}$. If the decay is  exponential $\delta
\rho(x,0) \sim e^{- x / r}$, then  $x_-(t) \sim r \log t$. These
are  the simple consequences of the Riemann  equation (\ref{H}).

The amplitude of  oscillations  is  zero at the trailing edge and
grows towards the leading edge. Furthermore, the period of the
oscillations also grows.  As a result, near the leading edge the
oscillatory pattern  resembles a collection of individual
localized traveling pulses (solitons) - a soliton train. The
excess velocity, $V-\frac{\kappa}{2}\overline{\delta\rho}$, of the
leading soliton is $\kappa \delta\rho$, which also determines  the
amplitude $4 \delta \rho$. The hight of the leading pulse   is
four times higher than the initial bump. Since each pulse carries
a quantized charge $1/\lambda$, the width of leading pulses, $(4
\pi \delta \overline{\delta\rho} \lambda)^{-1}$ stays constant,
while the distance between them is growing.

\paragraph{\it 15. Observation of shock waves and direct measurement
 of a fractional charge.}
Edge states of FQHE seems to be the best electronic  system to
observe quantum shock waves. In order to evaluate the scales one
needs parameters of the linearized  theory of the edge - the sound
velocity and the compressibility of the chiral boson. Although
these quantities have never been measured directly, one can
suggest reasonable  estimates \cite{Kang}. We estimate the sound
velocity on the edge as $v_s\sim 3\times 10^4$m/s. \cite{1998-MG} An electronic
density in the bulk $10^{11}$cm$^{-2}$ gives an estimate for the
1D density on the edge $\rho_0 \sim 3\times 10^7$m$^{-1}$. It
gives  $\kappa=v_{s}/\rho_{0} \sim 10^{-3}$m$^{2}$/s, and, at
$\lambda=3$ estimates an effective mass $m$ to be  about 30
electron band masses in GaAs and the ``Fermi'' time
$\tau_{F}=(\kappa\rho_{0}^{2})^{-1}$ to be $\sim 1$ps.

A wave-packet of  width $l\!\sim\! 10^2\rho_0^{-1}\!\sim\! 1\mu$m
and of  height $\delta\rho\!\sim\! 10^{-\!1}\rho_0$ carrying about
10 electrons, develops a shock wave at time $t_{c}\!\sim\! \tau_F
(\rho_0 l)\frac{\rho_0}{\delta\rho}\!\sim\! 1$ns. During the time
$t_{c}$ the wave packet crosses a distance $v_{s}t_{c}\!\sim\!
30\mu$m. This scale is much smaller than a size of a typical
heterostructure ($\!\sim\!10^3\mu$m), and is still smaller than
the typical ballistic length $50\!-\!100\mu$m. Distinct solitons
will start to appear right after  the shock. Observation of the
electric  charge carried by a distinct soliton-pulse will provide
a direct measurement  of fractional charge. The full decay of this
packet is about $10^3$ times longer.

These estimates suggest that  non-linear effects (shock waves and
fractionally charged soliton train)  can be observed in the
nanosecond range. Finally, we mention that quantum shock-waves in
Bose systems have already been observed in trapped alkali atoms
\cite{atoms}.

We have  benefited from discussions with  O. Agam, I. Krichever,
B. Spivak, A. Zabrodin, V. Goldman, and W. Kang. Our special
thanks to I. Gruzberg, D. Gutman, R. Teodorescu. P.W.  and E.B.
were supported by NSF MRSEC DMR-0213745 and NSF DMR-0220198. AGA
was supported by  NSF  DMR-0348358.


\begin{widetext}
\mbox{}
\end{widetext}

\textbf{ADDENDUM.} In this Addendum we provide some details of the
quantum Double Benjamin-Ono equation and its classical solution. A
fuller account will be published in \cite{paper5}.

\paragraph{\it 1. Holomorphic Bose field.}
The seemingly complicated equations, (\ref{5},\ref{6},\ref{7}),
possess integrable structure, following from  the integrability of
the Calogero model. This structure becomes more transparent  in
terms a anti-Hermitian \cite{f} holomorphic Bose field
$\varphi(z)$ and its current $ {\rm
u}=-i\frac{\kappa}{2\pi}\nabla\varphi(z)$ defined in the complex
plane $z=e^{i\frac{2\pi }{L}x}$ \cite{AW-2005}. The Bose field is
a Laurent series with respect to the unit circle $\varphi=
\varphi_+(z) + \varphi_-(z)$ having mode expansion outside of the
unit circle
\begin{equation}
 \nonumber
    \varphi_+(z)\!=\!\frac{1}{\lambda}\sum_{n> 0}
    \frac{1}{n}{\rm v}_n z^n,\;\;\varphi_-
    \!=\!\sum_{n<0}\frac{1}{n}\rho_{n} z^{n} + \frac{N}{2}\log z,\;\;
\end{equation}
$$[\rho_n,{\rm v}_{-m}]=n\delta_{nm}.
$$
Corresponding  currents are $ {\rm u}_-=-i\frac{\kappa}{2\pi}\nabla\varphi_-,\;
{\rm u}_+=-i\frac{\kappa}{2\pi}\nabla\varphi_+$.

The current is defined as a holomorphic operator having boundary
values  on the circle
\begin{eqnarray}
    {\rm u}_+(x) &=&  v -i\frac{\kappa}{2}\rho_H + i a\frac{\kappa }{4 \pi}\p_x\log\rho,
 \la{upb} \\
    {\rm u}_-(x\pm i0) &=& \frac{\kappa}{2}(i\rho_H\mp \rho).
 \la{umb}
\end{eqnarray}
Its jump is the proportional to the density $ {\rm
u}(x+i0)-{\rm u}(x-i0)=-\kappa\rho(x)$, thus defining the modes of the negative
part of the Laurent series through the moments of coordinates
$\rho_n=\oint e^{-i\frac{2\pi }{L}x n}\rho(x)dx $.

Let $ {\rm u}_+(z)$ and $ {\rm u}_-(z)$ be positive and negative
parts of the Laurent series for the current.  We have $ {\rm u }_-
= \frac{\kappa}{2} \oint \frac{dx}{L}\,\rho(x)\frac{ z + e^{i\frac{2\pi}{L}x }
 }{ z - e^{i\frac{2\pi}{L}x }} $ . This field $ {\rm
u}_-(z)$ is analytic inside and outside the unit disk, having  a
jump on its boundary   equal to the density.
The fact that the jump is real implies a reflection property
\be
    (\varphi_-(z))^\dag=\varphi_-(1/\bar z).
\ee

The field $u_+(z)$  is an analytic field in some strip
$R_1<|z|<R_2$ surrounding the circle, having value $u_+(x)= v -i
\frac{\kappa}{2}\rho_H + i a\frac{\kappa }{4
\pi}\p_x\log\rho$ on the circle. Under the chiral condition it
becomes $u_+(x)=\frac{\kappa}{2}(\rho - i \rho^H) + i a\frac{\kappa }{4
\pi}\nabla(\log\rho - i (\log \rho)^H)$. In this case
$u_+(z)$ becomes  analytic inside the circle (i.e., $R_1=0$
in the chiral case).

\paragraph{\it 2. Quantum bilinear equation}
In terms of the Bose field the equations of quantum hydrodynamics
are written in a compact way\cite{AW-2005}. Let us introduce
positive/negative vertex operators $\tau_+= e^{-\varphi_+}$ and
$\tau_-= e^{\lambda\varphi_-}$.
 Then eqs. (\ref{5},\ref{6}) are equivalently written
in the form of quantum  bilinear  Hirota equation
\begin{equation}\la{14}
(iD_t - \frac{\hbar }{2m} D_x^2)\,\tau_-\cdot\tau_+=0
\end{equation}
with the reflection condition
\begin{equation}\label{R}
\quad z^{- \lambda (2N -L\Phi )}(
\tau_-(1/\bar{z}))^\dag=\tau_-(z),
\end{equation}
where $\Phi$ is some constant. Hirota's derivatives are defined as
$D_x^n f\cdot g=(\p_x-\p_{x'})^nf(x)g(x')|_{x=x'}$. The bilinear
equation is a bosonized version of  Calogero  model.
 Its derivation from the original model and the derivation of
the quantum hydrodynamic eqs. (\ref{5}-\ref{7}) from the bilinear
equation is straightforward, but not effortless. Having a solution
of these equations one finds the density and velocity $ \rho =
\frac{1 }{ \lambda\pi }  \nabla \log \tau_-$, $v = \frac{ \kappa
}{ 2}\frac{1} {\lambda\pi } {\rm Im} \nabla \log \tau_+$. The
classical version of (\ref{14}), not bounded by a reflection
condition, is known as MKP (modified Kadomtsev-Petviashvili)
equation. Here it appears in a quantum version.

\paragraph{\it 3. Reductions of MKP: Benjamin-Ono and Double Benjamin-Ono equations.}
The bilinear form is the most economical way to understand
connections between eqs. (\ref{5},\ref{6},\ref{ch},\ref{BO}):
\begin{itemize}
\item[-]
The reflection
condition (\ref{R}) implies a reduction of MKP equation to the Double
 Benjamin-Ono equation (\ref{5},\ref{6}).

\item[-] Further requirement that $\tau_+$ is an analytical operator inside the unit disk
implies the chiral conditions leading to eq. (\ref{ch}).

\item[-] Finally further  reduction $u_-(z) =
\overline{u_+(1/\bar{z})} + C$, where $C$ is a real constant,
leads to the Benjamin-Ono equation
(\ref{BO}) \cite{1979-SatsumaIshimori}.
\end{itemize}

We publish a detailed study of this equation elsewhere \cite{paper5}. Here
 we present the
few major facts.

\paragraph{\it 4. Semi-classical limit}
The classical limit is straightforward in the bilinear form of the
quantum equation. There we simply treat $\tau_+$ and $\tau_-$ as
classical fields satisfying (\ref{14}) and (\ref{R}).

 Let us set $u_{\pm}=\mp \frac{\hbar}{m} i
\partial_x \log(\tau_\pm)$,
then  the classical  bilinear equation  (\ref{14}) reads
$$\dot u+u\nabla u+\frac{\hbar}{2m}\nabla^2 \tilde u=0$$
where $u=u_++u_-$ and $\tilde u=i(u_--u_+)$.

\paragraph {\it 5 Multiphase solution}
A general quasi-periodic (or $N$-multiphase) solution of classical
MKP equation (\ref{5},\ref{6}) (in the frame where the center of
mass of the system is at rest) is coded by $2N$  parameters - $
v_i,\tilde v_i$ (wave number $k_i=m(v_i-\tilde v_i)$ and
velocities $V_i = \frac{1 }{ 2 } (v_i+\tilde v_i) + v_0 $ of
waves), $N$ real phases $x_i$, $2N$ real positive amplitudes
 $c_i^{\pm}$ and two  zero modes $K$ and $v_0$. It is given by a
determinant formula for the tau-functions
\begin{equation} \label{det}
\tau_\pm(x+i0,t)=e^{ \frac{i}{\hbar} \theta_\pm}\det_{i,j\leq N}
\left( \delta_{ij} -  c_i^{\pm}  \frac{
e^{\frac{i}{\hbar}\theta_i(x,t)}}{ v_i - \tilde v_j }\right).
\end{equation}
Here $\theta_i(x,t)\!= \! k_i(x-x_i - V_i  t)$  and $ \theta_- =-
m K (x-\frac{K}{2}t)$, $ \theta_+ = m v_0 (x- (\frac{1}{2} v_0+K)
t) $, where $k_i = v_i - \tilde{v}_i $, and $V_i = \frac{
\tilde{v}_i + v_i } {2} + v_0$.


The reflection property, (\ref{R}), further restricts the
amplitudes
\begin{eqnarray} \nonumber
(c_i^{(\pm)})^2\!= \frac{ k_i^2}{m^2}\left(\!\frac{ v_i \! - \!K}
{\tilde v_i \! -\! K \!} \! \right)^{ \pm 1 }\!\left( \! \frac{
 v_i  + \!K\! - \!\Phi }{\tilde v_i  + \! K \! - \! \Phi }
\!\right) \frac {\prod_{j \neq i }(\tilde v_j \!  - \! v_i )( v_j
\! - \! \tilde v_i)}{\prod_{j\neq i}(v_j\!-\!v_i)(\tilde
v_j\!-\!\tilde v_i)},
\end{eqnarray}
where  $\Phi=\frac{1}{m}\sum_i k_i$. The parameter $\Phi$
determines monodromy of the wave function. It is the flux piercing
the circle. Additional restrictions on $v_i,\tilde v_i$  insure
that $c_i$  are real. The zero mode $K$ is related to the mean
density $ \rho_0 =  \frac{ 2}{\kappa}( \Phi + K)$. The mean
velocity is $v_0$.

 One-phase solution was found in
\cite{Polychronakos-1995}.

The chiral case occurs when all zeros of  $\tau_-$  are located
outside  the unit disk of the complex plane  $z=e^{i\frac{2\pi}{L}
x}$,
 while all zeros  and those of $\tau_+$ are inside the unit
disk.   \indent Solutions of the BO eq. (\ref{BO}) arise at the
limit $ \frac{v_i-K}{K} \to 0$. This yields (\ref{det}) with the
amplitudes \cite{1979-SatsumaIshimori}
\begin{eqnarray} \nonumber
    (c_i^{(\pm)})^2\!= \frac{ k_i^2}{m^2}\left(\!\frac{ v_i -K } {\tilde
    v_i - K } \! \right)^{ \pm 1 } \frac {\prod_{j \neq i }( \tilde v_j\! -
    \! v_i )( v_j \! - \! \tilde v_i)}{\prod_{j\neq
    i}(v_j\!-\!v_i)(\tilde v_j\!-\!\tilde v_i)}.
\end{eqnarray}

\paragraph{
6.  Multisoliton solutions} of the DBO equations (\ref{5},\ref{6})
appear as a long-wave limit  of (\ref{det}), obtained by taking
$k_i\to 0$, while keeping $V_i$ fixed:
\begin{equation}\nonumber
    \tau_\pm\!=\!e^{ -\frac{i}{\hbar} \theta_\pm }\!\det \left(
    (x\!-\!x_i\!-\!V_it\!-\!iA_i^{\pm} ) \delta_{ ij }\! -   \!
    \frac{i \hbar }{m} \frac{ 1\!-\!\delta_{ij}}{ V_i\!-\!V_j }
    \right),
\end{equation}
where $ A_i^{\pm} =  \frac{\hbar}{2m} \left( \frac{1}{ V_i + K  }
 \pm \frac{1}{ V_i - K  } \right)$.
  A limit of this equation $(V_i-K)/K\to 0$  gives the
multisoliton solution of the Benjamin-Ono equation $A_i^{\pm} =
\pm \frac{\hbar }{2 m (V_i-K)}$ \cite{1979-SatsumaIshimori}.

\begin{thebibliography}{19}
\bibitem{f}The hydrodynamic operators act in the space of symmetric functions of coordinates with  the inner product $\langle f,g\rangle=\oint d x_i|\Delta(x)|^{2\lambda}
f(x)g(x)$, where $\Delta^\lambda(x_i) =\prod_{i\neq
j}(\sin(\frac{\pi}{L}(x_i-x_j)))^\lambda$ is
the ground state  of (\ref{1}).
\end{thebibliography}
\end{document}